\documentclass[%
 reprint,
superscriptaddress,
 amsmath,amssymb,
 aps,
floatfix,
]{revtex4-2}

\usepackage{graphicx}
\usepackage{dcolumn}
\usepackage{bm}
\usepackage[colorlinks=true, allcolors=blue]{hyperref}

\usepackage{dsfont}
\usepackage{xcolor}
\usepackage{todonotes}
\usepackage{ulem}
\usepackage{listings}
\usepackage{physics}
\usepackage{float}
\usepackage{siunitx}

\usepackage{caption}[justification=justified]
\usepackage{subcaption}[justification=justified]

\begin{document}

\preprint{APS/123-QED}

\title{Modular source for near-infrared quantum communication}

\author{Federico~Berra}
\thanks{These authors contributed equally to this work.}
\affiliation{%
 Dipartimento di Ingegneria dell'Informazione, Universit\`a degli Studi di Padova, via Gradenigo 6B, 35131 Padova, Italy
}%

\author{Costantino~Agnesi}
\thanks{These authors contributed equally to this work.}
\affiliation{%
 Dipartimento di Ingegneria dell'Informazione, Universit\`a degli Studi di Padova, via Gradenigo 6B, 35131 Padova, Italy
}%

\author{Andrea~Stanco}
\affiliation{%
 Dipartimento di Ingegneria dell'Informazione, Universit\`a degli Studi di Padova, via Gradenigo 6B, 35131 Padova, Italy
}%

\author{Marco~Avesani}
\affiliation{%
 Dipartimento di Ingegneria dell'Informazione, Universit\`a degli Studi di Padova, via Gradenigo 6B, 35131 Padova, Italy
}%

\author{Sebastiano~Cocchi}
\affiliation{%
 Dipartimento di Ingegneria dell'Informazione, Universit\`a degli Studi di Padova, via Gradenigo 6B, 35131 Padova, Italy
}%

\author{Paolo~Villoresi}
\affiliation{%
 Dipartimento di Ingegneria dell'Informazione, Universit\`a degli Studi di Padova, via Gradenigo 6B, 35131 Padova, Italy
}%
\affiliation{%
 Padua Quantum Technologies Research Center, Universit\`a degli Studi di Padova, via Gradenigo 6B, 35131 Padova, Italy
}%

\author{Giuseppe~Vallone}
\email{giuseppe.vallone@unipd.it}
\affiliation{%
 Dipartimento di Ingegneria dell'Informazione, Universit\`a degli Studi di Padova, via Gradenigo 6B, 35131 Padova, Italy
}%
\affiliation{%
 Padua Quantum Technologies Research Center, Universit\`a degli Studi di Padova, via Gradenigo 6B, 35131 Padova, Italy
}%
\affiliation{%
 Dipartimento di Fisica e Astronomia, Università degli Studi di Padova, via Marzolo 8, 35131 Padova, Italy
}%

\date{\today}

\begin{abstract}
We present a source of states for Quantum Key Distribution (QKD) based on a modular design exploiting the iPOGNAC, a stable, low-error, and calibration-free polarization modulation scheme, for both intensity 
and polarization encoding. This source is immune to the security vulnerabilities of other state sources such as side channels and some quantum hacking attacks. Furthermore, our intensity modulation scheme allows full tunability of the intensity ratio between the decoy and signal states, and mitigates patterning effects. The source was implemented and tested at the near-infrared optical band around 800 nm, of particular interest for satellite-based QKD. Remarkably, the modularity of the source simplifies its development, testing, and qualification, especially for space missions. For these reasons, our work paves the way for the development of the second generation of QKD satellites that can guarantee excellent performances at higher security levels. 
\end{abstract}

\maketitle
\section{Introduction}
Quantum Key Distribution (QKD)~\cite{Gisin2002, Pirandola2019rev} is essential to ensure the safe exchange of sensitive data between distant parties.
Establishing its security on the principles of quantum mechanics and the characteristics of photons, QKD allows two distant parties to distill a secret key with unconditionally secure and bound the shared information with any adversarial eavesdropper~\cite{Scarani2008}.
Furthermore, unlike computationally-secure classical algorithms, QKD offers long-term privacy since algorithmic and technological advances for both classical and quantum computation do not threaten the security of keys generated with QKD. 

Satellite-based QKD~\cite{Agnesi2018, Kaltenbaek2021, Sidhu2021} is essential for the development of a global-scale network mainly because the achievable distance between parties with a satellite-assisted link is substantially larger than the distances compatible with optical fiber which is limited by exponential propagation losses to a few hundred of kilometers~\cite{Boaron2018} in the absence of quantum repeaters.
This has led to several pioneering works in satellite quantum communications~\cite{Villoresi_2008,Vallone2015,Vallone2016}, culminating in the development and launch of the Micius satellite by the Chinese Academy of Science~\cite{Liao2017_Sat} that demonstrated intercontinental QKD links~\cite{Liao2018}. In this regard, the near-infrared (NIR) optical band around 800 nm  has been often cited as an ideal wavelength for satellite-based quantum communications since it has good atmospheric transmission, enables the use of free-space coupled silicon-based single photon avalanche diode (SPADs), and is a good compromise in terms of beam divergence (which is proportional to the wavelength) especially when compared to longer wavelengths.

The technical solution employed by the Micius satellite to develop the QKD transmitter is based on a multiple light source approach, where each polarization state and each intensity state was emitted by an independent laser. This leads to a total of 8 lasers being used to implement the decoy-states BB84  protocol~\cite{Bennett2014_BB84, Hwang2003}. This solution offers good performances in terms of stability and intrinsic QBER,  but recent studies have highlighted that a fully secure implementation can be challenging~\cite{Pirandola2019rev, Lo2014}.

A first concern is related to the distinguishability of the optical pulses emitted by the independent laser sources and responsible for encoding the different polarization and intensity states. Any difference between the photonic degrees of freedom of the light pulses, such as in the spectral or temporal profiles, could enable an eavesdropper to perform a side-channel attack, obtaining information about the exchanged key without being detected and compromising the security of the protocol~\cite{Nauerth_2009}. If not properly assessed and mitigated, the harsh space environment could exacerbate this security vulnerability since each individual laser could be subject to different temperature gradients or radiation doses, individually modifying their behavior and opening a side channel for a quantum hacker to exploit. The second vulnerability of the multiple light source approach is that it is susceptible to  some quantum hacking attacks such as the Trojan Horse attack described by Lee \textit{et al.}~\cite{Lee2019}, where an eavesdropper can change the wavelength of the independent laser sources of different amounts, enabling him to obtain polarization information without performing a direct polarization measurement.

A possible solution to these security concerns is to change the design of the QKD transmitter to implement decoy-states BB84 with a single light source, an intensity modulator to generate the decoys, and a polarization modulator to encode the quantum states.
This, however, comes with the technical challenge of developing intensity and polarization modulation stages that guarantee the required performances in terms of stability and state quality. Regarding, intensity modulation a large concern emerged with the patterning effect that commercial-off-the-shelf intensity modulators would exhibit and would cause a significant decrease in the achievable secure key rate~\cite{Yoshino2018}. However, this patterning effect was mitigated with the design presented by Roberts \textit{et al.}~\cite{Roberts2018} at the cost of fixing the decoy state ratio at construction.
Regarding polarization modulation instead, the iPOGNAC~\footnote{The iPOGNAC is object of the Italian Patent No. 102019000019373 filed on 21.10.2019 as well as of the \href{https://patentscope.wipo.int/search/en/detail.jsf?docId=WO2021078723}{International Patent Application no. PCT/EP2020/079471} filed on 20.10.2020.} offers a stable, low-error, and calibration-free solution~\cite{Avesani2020}, which has currently been developed and tested only at 1550 nm.

In this work, we present a novel QKD source designed for satellite-based operations and working in the NIR optical band around 800 nm. This QKD source  adopts a modular design approach, exploiting the iPOGNAC for both intensity and polarization modulation. In this way, patterning-effect-free intensity modulation is obtained with the added flexibility of effortless tuning the intensity ratio. Furthermore, polarization modulation with the iPOGNAC guarantees polarization states that are fixed with respect to the transmitter’s reference frame eliminates the need of calibration between the transmitter
and the receiver. Secondly, given its free-space output, it can
be easily interfaced with a telescope, making it a promising solution for
quantum communication with satellites. 

The manuscript is organized as follows: The design and working principle of our modular QKD source are explained in Section~\ref{sec:Setup}, giving particular focus to the novel iPOGNAC-based intensity modulation scheme. Experimental validation of the source is performed in Section~\ref{sec:results} that concludes with a proof-of-principle QKD experiment.

\section{Setup}
\label{sec:Setup}

\subsection{Intensity Modulation}
\begin{figure}[h]
  \includegraphics[width=\linewidth]{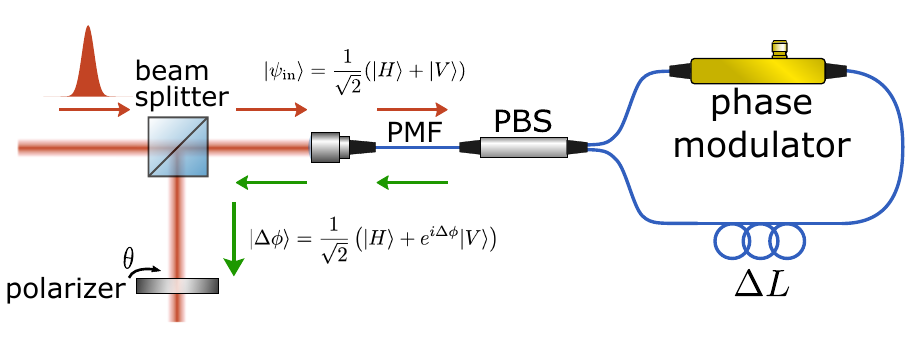}
  \caption{Scheme of the proposed intensity modulator composed of an iPOGNAC polarization modulator followed by a polarizer rotated at an angle $\theta$  .
  }
  \label{fig:intensity}
\end{figure}
\label{section:IntModulator}
\begin{figure*}
\includegraphics[width=\textwidth]{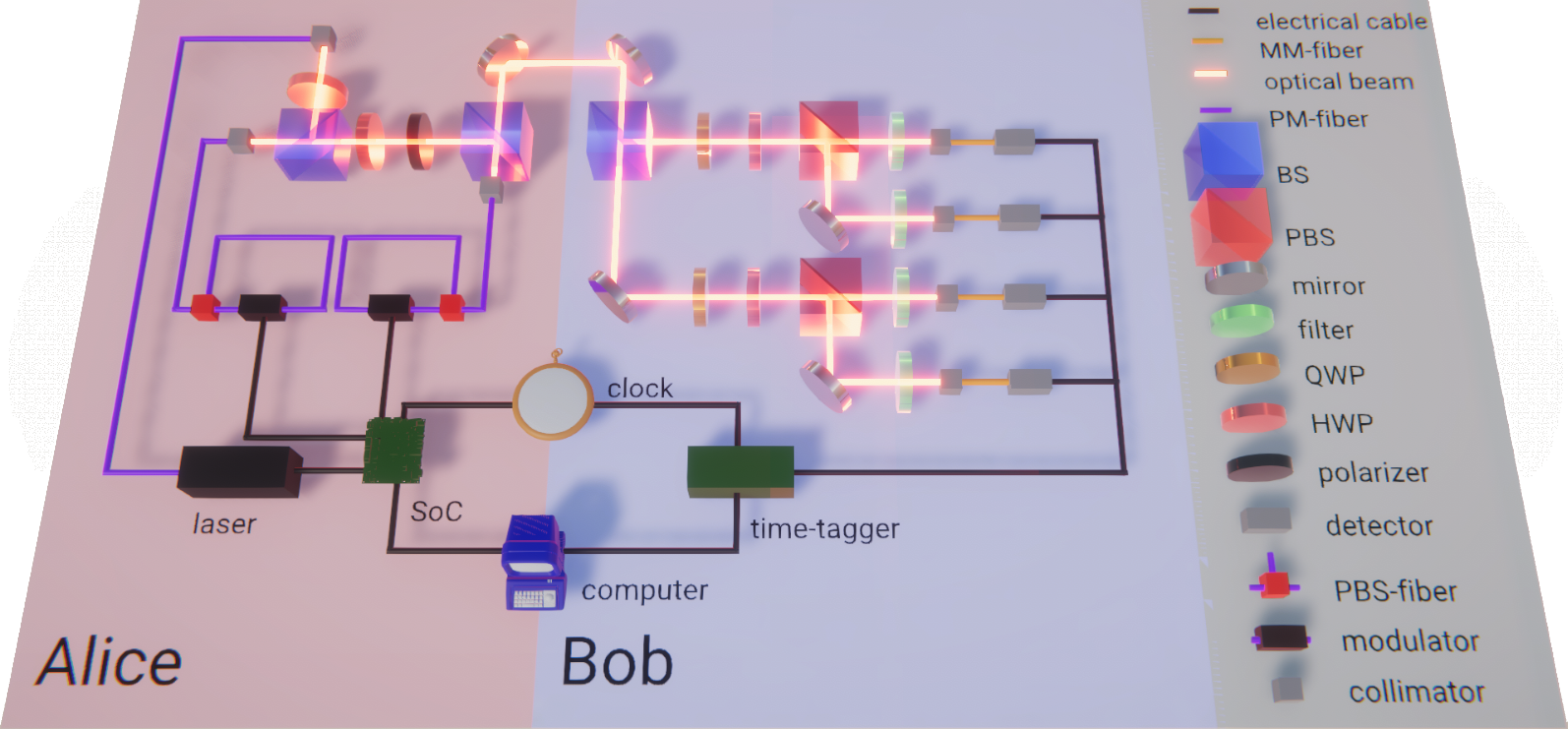}
  \caption{A rendered version of the modular source for near-infrared quantum communications. }
  \label{fig:optical-setup}
\end{figure*}

The intensity modulator introduced in this work, depicted in Fig.~\ref{fig:intensity}, is based on the iPOGNAC polarization modulator~\cite{Avesani2020}. This design choice results in our intensity modulator inheriting all of the key performance characteristics of the iPOGNAC.
In particular, its self-compensating design leads to long-term stability without the need for any feedback mechanism.
This has been thoroughly tested in previous works~\cite{Avesani2020, Scalcon22}, even in an urban field trail~\cite{Avesani:21}.  Furthermore, compared to other polarization encoders, the iPOGNAC is capable of producing fixed, stable, and well-defined polarization states without any need for calibration. This fact is exploited in the construction of the intensity modulator.

To achieve these characteristics, the iPOGNAC combines a hybrid free-space and fiber-optical scheme, obtaining the polarization stability of free-space optics as well as the flexibility and technological maturity of fiber-based optical components. The iPOGNAC begins with a free-space segment composed of a half-wave plate (HWP) and a beamsplitter (BS). The HWP is used to convert the input linearly polarized light pulses to a diagonal state of polarization (SOP) $\ket{D} = \left( \ket{H} + \ket{V} \right ) / \sqrt{2}$. Instead, the BS is used to separate the input beam from the output. The light is then coupled into a polarization-maintaining (PM) optical fiber and sent to an unbalanced Sagnac interferometer containing a high-bandwidth phase modulator. Here, however, the BS is replaced by a fiber-based polarization beamsplitter (PBS) with a PM optical fiber input and outputs.
The asymmetry of the interferometer allows us to control the SOP exiting the device by properly setting the voltage and the timing of the pulses driving the phase modulator as follows:
\begin{equation}
    \ket{\Delta\phi} = \frac{1}{\sqrt{2}} \left ( \ket{H}  + e^{i  \Delta\phi} \ket{V} \right )
    \label{eq:iPOGNACstate}
\end{equation}
where $\Delta\phi = \phi_{\mathrm{CW}} - \phi_{\mathrm{CCW}}$, and $\phi_{\mathrm{CW}}$ and $\phi_{\mathrm{CCW}}$ are  the phases applied by the phase modulator to the clockwise (CW) and counter-clockwise (CCW) propagating light pulses respectively.
In particular, if we apply a voltage pulse that induces a $\pi$ phase shift to either the CW or the CCW light pulses, the iPOGNAC generates the antidiagonal SOP $\ket{A} = \left( \ket{H} - \ket{V} \right ) / \sqrt{2} $. Instead, if no phase shifts are applied, the SOP remains $\ket{D}$. 
These two states are fundamental in the operation of our iPOGNAC-based intensity modulator, since we target modulating between two mean photon number levels, as required for the 1-decoy state QKD protocol~\cite{Rusca2018_APL}. This decoy-state scheme is chosen as it simplifies the requirements of
the quantum state encoder and can provide higher rates in the
finite-key scenario~\cite{Rusca2018_APL}. 
The light pulses then travel back through the PM fiber and are emitted onto the free-space once again, where the BS directs the light toward the free-space output port.

What distinguishes our intensity modulator from a standard iPOGNAC polarization modulator is that we place a polarizer, with a rotation angle $\theta$, at the output port.

The polarizer rotated at an angle $\theta$ results in a projection onto the state $\ket{\theta} = \cos(\theta) \ket{H} + \sin(\theta) \ket{V}$ which can be rewritten as $\ket{\theta} = \cos(\theta - \frac{\pi}{4}) \ket{D} + \sin(\theta - \frac{\pi}{4}) \ket{A}$ to simplify calculations. When the $\ket{D}$ SOP encounters the polarizer, its transmission probability is given by 
$ | \braket{\theta}{D} |^2 = \cos^2(\theta - \pi/4) $,whereas the transmission probability for the $\ket{A}$ state is given by $
   | \braket{\theta}{A} |^2 = \sin^2(\theta - \pi/4)$. 
From this, we obtain the intensity  ratio value between these two possible states is given by:
\begin{equation}
    \label{eq:decoyER}
    \mathrm{IR}(\theta) = \frac{| \braket{\theta}{A} |^2}{| \braket{\theta}{D} |^2}  = \tan^2 \left ( \theta - \frac{\pi}{4} \right ).
\end{equation}

From Eq.~\ref{eq:decoyER}, it is clear that the intensity ratio between the two states can be easily tuned to any value by changing the polarizer angle $\theta$, with physical device imperfections representing the only limit. 
This feature makes our iPOGNAC-based intensity modulator more flexible than other self-compensating intensity modulators such as the one introduced by Roberts \textit{et al.}~\cite{Roberts2018}, which has an intensity ratio that is fixed at construction by the transmissivity and reflectivity of the beam splitter used in their Sagnac interferometer.
Tuning this ratio can be crucial to obtain the best performance of the QKD system since a change to the operational scenario could lead to a different optimal setting for the decoy states~\cite{Rusca2018_APL}. Furthermore, this feature simplifies the construction and industrialization of the intensity modulator since its performance is not dependent on the fabrication tolerances of the optical components, leading to higher standards of quality and performance repeatability. 

\begin{figure}[b]
  \includegraphics[width=\linewidth]{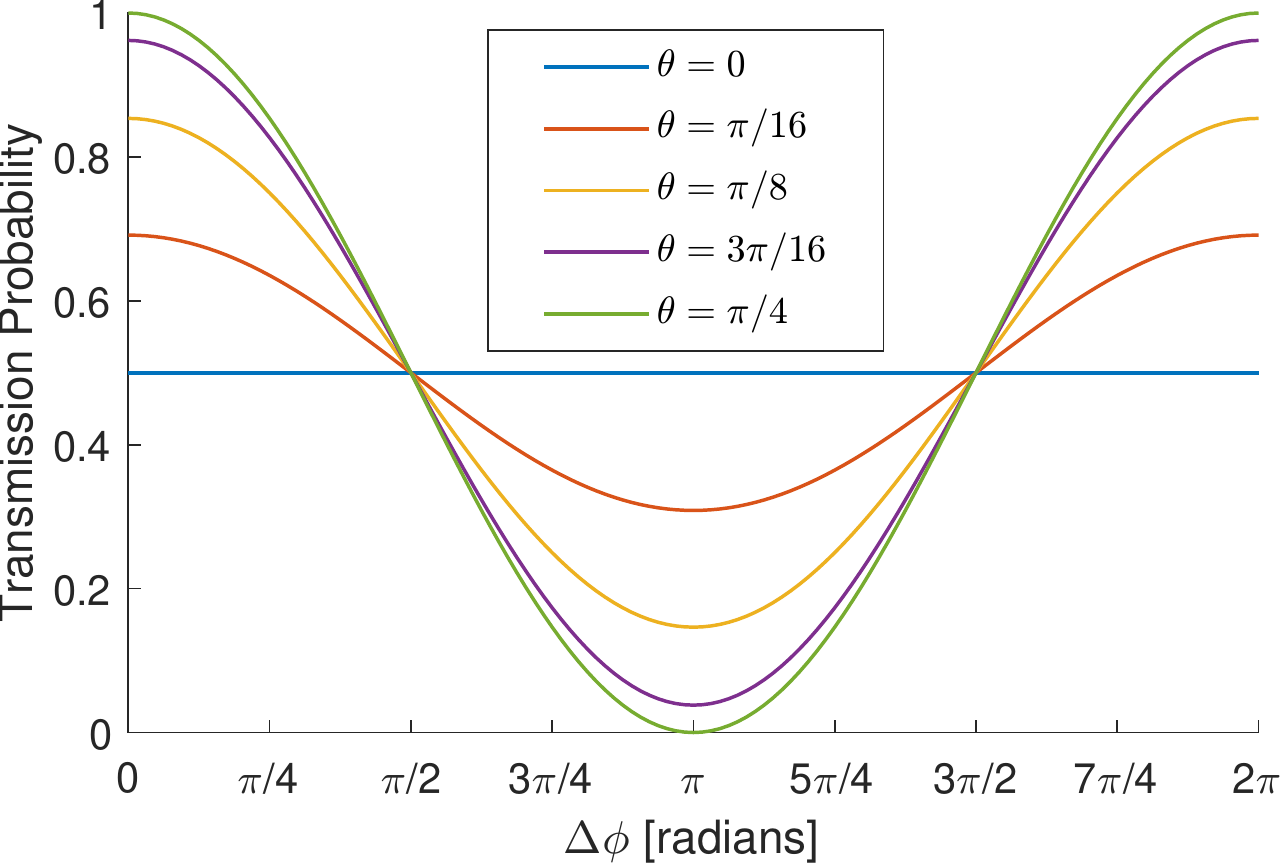}
  \caption{Theoretical optical response of the system as a function of the iPOGNAC polarization phase modulation $\Delta\phi$ (see Eq.~\ref{eq:iPOGNACstate}) for different polarizer rotation angles $\theta$.}
  \label{fig:intTheory}
\end{figure}

Another key feature of our intensity modulator is that it is free from the patterning effect. This effect arises when the intensity of a pulse emitted by the transmitter depends on the previous pulse intensity. This is a security concern for the implementation of the decoy-state method and results in a significant drop in the achievable secure rate when taken into account~\cite{Yoshino2018}. The patterning effect can be mitigated by working at the points with vanishing derivative of the optical response function~\cite{Roberts2018} since in the latter points, small deviations caused by imperfections and the finite modulation bandwidth of the system cause only small variations in the intensity ratio. In our design, this is guaranteed by using orthogonal SOPs $\ket{D}$ ($\Delta\phi=0$) and $\ket{A}$ ($\Delta\phi=\pi$), which always correspond to the peak and trough points of the optical response function for all values of the polarizer angle $\theta$, as inferred from Fig.~\ref{fig:intTheory}.
Similar patterning-effect mitigation could have been obtained by applying $\pi/2$ radians phase shifts and obtaining the orthogonal circular left $\ket{L} = \left( \ket{H} + i \ket{V} \right ) / \sqrt{2} $ and circular right $\ket{R} = \left( \ket{H} - i \ket{V} \right ) / \sqrt{2}$ SOPs. 
However, this would have increased the complexity of the setup since a quarter-wave plate (QWP) would have been introduced to perform the required projection and coordinated rotation of the QWP and the polarizer would have been necessary. We also note that at a fixed polarizer angle, by changing the value of $\Delta \phi$ any intensity ratio between 0 and the value predicted by Eq.~\ref{eq:decoyER} can be obtained. Therefore, different intensity levels can be generated by using different values of $\Delta \phi$, but a patterning effect might emerge.

\subsection{Modular QKD source}
We developed a QKD source capable of implementing efficient 3-states 1-decoy BB84 protocol~\cite{Grunenfelder2018} working in the NIR optical band.
The light source used at the transmitter is a gain-switched  PM fiber-coupled distributed-feedback laser (Eagleyard EYP-DFB-0795), emitting $795$~nm light pulses with $575$~ps FWHM at a repetition rate of $R = 50$~MHz and driven by a laser pulser (Highland Technology T165). 
A PM fiber-based polarizer is then encountered to guarantee a stable and fixed SOP as the input for the iPOGNAC-based intensity modulator, described in detail in Section~\ref{section:IntModulator}. For convenience, instead of rotating the intensity modulator's polarizer, we decided to keep it at a fixed angle and inserted an HWP before it to emulate the polarization rotation angle. This allowed us to have a fixed output polarization state $\ket{D}$ at the output of the intensity modulator without changing the characteristics of the device and simplifying the interface with the following module.
The HWP was set at an equivalent polarizer angle $\theta \approx 0.50$ rad, tuned to guarantee a signal  and decoy  ratio of $\nu/\mu\approx 0.30$ which is near optimal for the three-state and one-decoy efficient BB84 protocol for a wide range of total losses (30~dB to  60~dB) of interest for satellite-based QKD~\cite{Rusca2018_APL}. 

The light then encountered a second iPOGNAC encoder, responsible for modulating the degree of freedom of polarization of the qubit. In this case, the driving electric pulse amplitude was set to induce a $\pi/2$ phase shift, allowing the iPOGNAC to generate circular left $\ket{L}$, circular right $\ket{R}$, or diagonal $\ket{D}$ polarized light.
In this way, we generate the three states required by the simplified three-polarization state version of BB84, with the key generation basis $\mathcal{Z} = \{\ket{0}, \ket{1}\}$ where $\ket{0}:=\ket{L}$, $\ket{1}:= \ket{R}$, and the control state $\ket{+}$ of the $\mathcal{X}=  \{\ket{+}, \ket{-}\}$ control basis where $\ket{+}:=\ket{D}$, $\ket{-}:= \ket{A}$. 
A Variable Optical Attenuator (VOA) then sets an appropriate intensity for signal ($\mu \approx 0.6$) and decoy ($\nu \approx 0.2$) pulses.
The light was then sent to the quantum receiver via a free-space channel.

The electronic signals that trigger the laser pulser and drive the modulators are controlled by a system-on-a-chip (SoC) that includes a field-programmable gate array (FPGA) and a CPU~\cite{Stanco2022} and is integrated on a dedicated board (Zedboard by Avnet).

\subsection{QKD Receiver}

The quantum state receiver is based on a well-tested and fully free-space design that has been used even in satellite-based QKD experiments~\cite{Liao2017_Sat}. The measurement basis choice is performed passively using a 60:40 BS. At each output port of the BS, QWPs, HWPs, and PBSs are placed to perform projective measurements. In particular, the transmitted light (60\%) is measured in the key-generation basis $\mathcal{Z}$, whereas the reflected light (40\%) is measured in the $\mathcal{X}$ control basis. After projection, light is filtered by 10 nm FWHM passband filters and collected by multimode fibers ( $NA = 0.22$ and $105~\mu m$ core size) which guide light toward silicon-based single-photon avalanche diodes (SPAD) with 68\% quantum efficiency and about 1000 dark counts per second. A time-to-digital converter was used to record the detection events that were then processed by a computer.

In our setup, synchronization between the transmitter and the receiver can be performed via a direct RF cable link, exploiting a clock-data-recovery routine performed on a co-propagating classical optical link~\cite{Berra2022_sync}, or via Qubit4Sync qubit-based synchronization~\cite{Calderaro2020}.
However, for experimental simplicity, a direct RF cable link was preferred. 

To simplify the transportation and installation of the receiver, particular attention was paid to reducing its footprint. Taking advantage of the vertical direction, the complete receiver was contained on a $0.3~m \times 0.3~m$  optical breadboard. In particular, this was achieved by orienting the reflected port of the projection PBSs upward.

\section{Results}
\label{sec:results}
\subsection{Tunability of the intensity}

As mentioned in Section~\ref{section:IntModulator}, the first key feature of the iPOGNAC-based intensity modulator is its capability of tuning the optimal ratio between the two intensity levels $\mu, \nu$ simply by rotating the polarizer at the end of the intensity modulator.

\begin{figure}[b]
  \includegraphics[width=\linewidth]{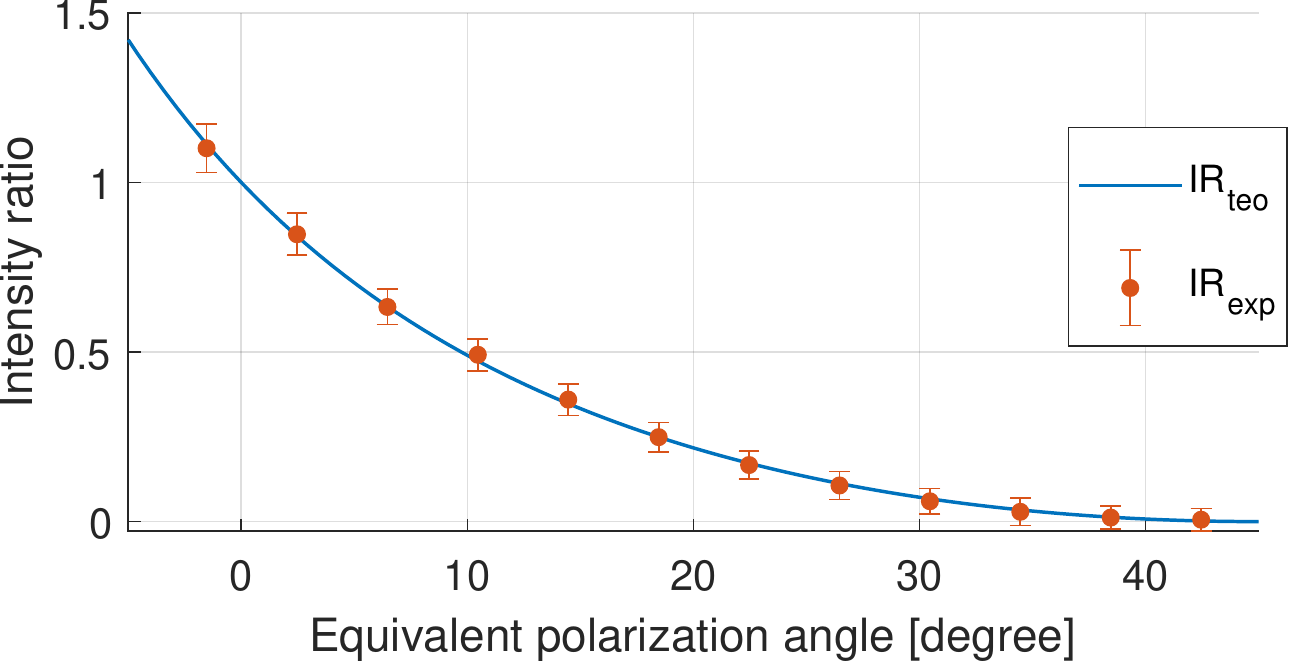}
  \caption{Ratio between $\nu$ and $\mu$ intensities: the dots represent the experimental data with associated error bars, whereas the continuous line is derived from Eq.~\ref{eq:decoyER}.} 
  \label{fig:malus-law}
\end{figure}
We tested this behavior using the setup described in the previous section, shown in Fig.\ref{fig:optical-setup}, by sending a pseudorandom sequence of intensities and tacking a $60~s$ acquisition for each equivalent polarizer angle obtained by rotating an HWP. As reported in Fig.\ref{fig:malus-law}, a total of 12 different equivalent polarizer angles were tested in the range around 0 and $\pi/4$, all in good correspondence with the theoretical values obtained from Eq.~\ref{eq:decoyER}.

\subsection{Patterning Effect Mitigation}
\begin{figure}
  \includegraphics[width=\linewidth]{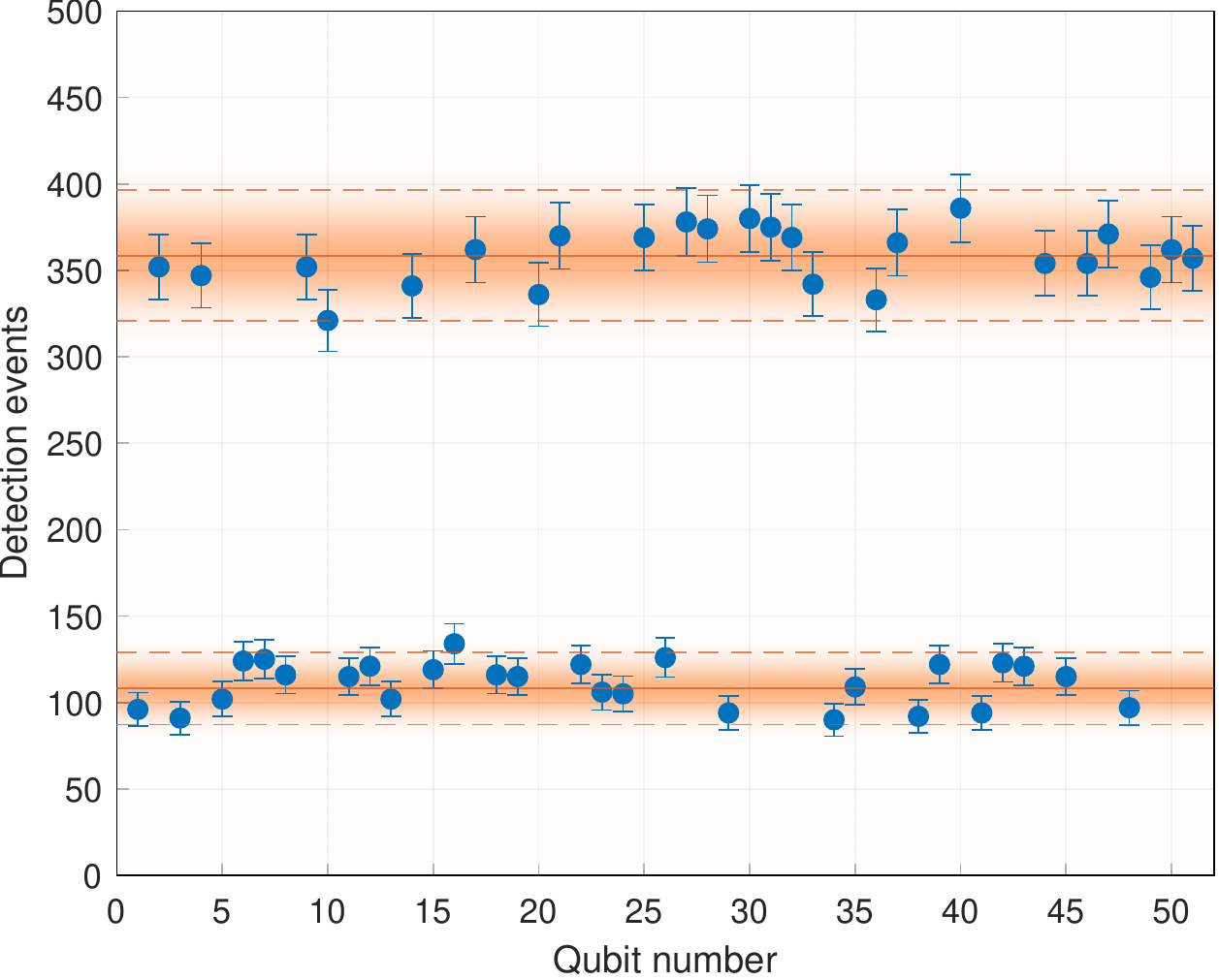}
  
  \caption{Single photon detection statistics (blue) for a $50$-symbol portion of a random $\mu, \nu$ intensity pattern: the dots represent the experimental data with associated error bars, instead, the  solid line represents the average value whereas the dashed lines represent the $\pm2\sigma$ confidence interval.}

  \label{fig:20-peaks}
\end{figure}

The second key feature of the intensity modulator, as explained in Section~\ref{section:IntModulator}, is the fact that it mitigates patterning effects by operating at the peak of the optical response function where the derivate is smaller. This guarantees that fluctuation of the driving electric signal produces small deviations in the final intensities.

As before, we tested this behavior using the setup described shown in Fig.\ref{fig:optical-setup} by sending a 1024-bit pseudorandom sequence of intensities and tacking a $120~s$ acquisition for a polarizer angle of $\theta \approx 0.50$ rad, tuned to guarantee a signal and decoy ratio of $\nu/\mu \approx 0.30$. The detection histogram for a subset of  50 intensities can be seen in Fig.~\ref{fig:20-peaks}.  

For each intensity, we computed the normalized average intensity of its subsequent pulse: 
\begin{equation}
    \label{eq:average-intensity}
    c_{i \to i'} = \frac{\langle s_{i \to i'} \rangle }{\langle\mu\rangle}
\end{equation}
and the deviation from the average:
\begin{equation}
    \label{eq:deviation-from-average}
    d_{i \to i'} = \frac{\langle s_{i \to i'} - \langle i'\rangle\rangle}{\langle i'\rangle}
\end{equation}
where $s_{i \to i'}$ is the click's count for the symbol $i'$ with preceding symbol $i$, and $\langle i'\rangle$ is the average between all the same symbols. The results reported in Tab.~\ref{tab:patterning} show that all the fluctuations are within the experimental uncertainty and confirm that there is no patterning. This result is a substantial improvement compared to the best-case scenario of around $18.2\%$ deviations observed by Yoshino \textit{et al.}.~\cite{Yoshino2018} when producing decoy states using a commercial Mach-Zehnder intensity modulator at the quadrature point, and is in line with the results obtained by Roberts \textit{et al.}~\cite{Roberts2018}.

\begin{table}
\centering
\begin{tabular}{ c | c |  c }
Pattern & $c_{i \to i'}$ & $d_{i \to i'}$ (\%) \\
\hline \hline
$\mu \to \mu$ & $1.00 \pm 0.04$ & $0.001$ \\
$\nu \to \mu$ & $1.00 \pm 0.04$ & $-0.001$ \\
$\nu \to \nu$ & $0.30 \pm 0.02$ & $-0.001$ \\
$\mu \to \nu$ & $0.30 \pm 0.02$ & $0.001$ \\   

\end{tabular}
\caption{Average pulse intensities of $\mu$ and $\nu$ when preceded by either $\mu$ or $\nu$. The average pulse intensity for the $\mu$ intensity is normalized to unity.  }
\label{tab:patterning}
\end{table}

\subsection{QKD Experiment}
\begin{figure}
  \includegraphics[width=\linewidth]{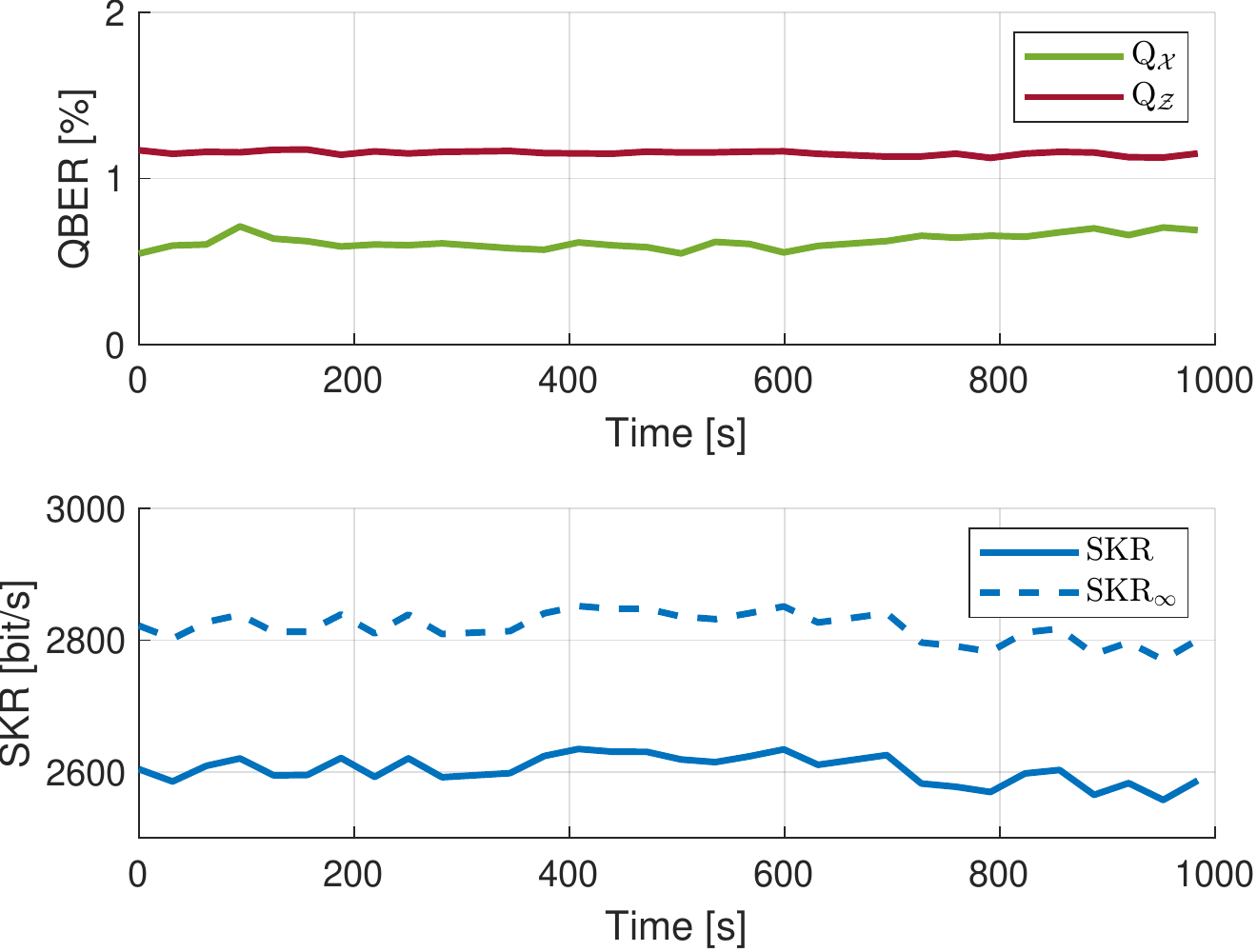}
  \caption{The Quantum Bit Error Rate and the Secure Key Rate obtained with our modular QKD source. An average $\mathrm{Q}_\mathcal{Z} = 0.62 \pm 0.05$ and $\mathrm{Q}_\mathcal{X} = 1.15 \pm 0.01$ were observed whereas a finite-key  $\mathrm{SKR} = 2603 \pm 21$ ($\mathrm{SKR}_\infty = 2819 \pm 23$) bits per second was obtained.   }
  \label{fig:qber-skr}
\end{figure}

To evaluate the overall performances of our modular quantum source, we performed a proof-of-principle 15-minute long QKD experiment.
Such a duration was targeted since it represents the typical duration of a Low Earth Orbit satellite passage~\cite{Liao2017_Sat}. 
The test was performed using a quantum channel consisting of a free-space segment and attenuating neutral density filters to simulate losses caused by geometrical losses and atmospheric absorption typical of satellite links. 
The mean detection rate $R_\mathrm{det}$ was of $\approx 2.7\cdot10^{5}$ events per seconds. 
Considering that on average the source emitted $(\mu P_\mu + \nu P_\nu)\cdot R  = 2.4\cdot10^{7}$ photons per second, the measured total losses were approximately 19 dB. 
The channel contribution to these losses is about 15 dB, while the remaining 4 dB can be attributed to the detectors' efficiencies and other receiver losses. 

We report the quantum bit error rate (QBER) and the secret key rate (SKR) obtained in Fig.~\ref{fig:qber-skr}. The QBER was calculated independently for the key generation basis $\mathcal{Z}$ and the control basis $\mathcal{X}$. We can see that both QBERs are lower than the $\approx 11$ \% upper limit for secure key generation, with $\mathrm{Q}_\mathcal{Z} = 0.62 \pm 0.05$ and $\mathrm{Q}_\mathcal{X} = 1.15 \pm 0.01$. The SKR was calculated following the finite-size analysis of Ref.~\cite{Rusca2018_APL}:
\begin{equation}
 \mathrm{SKR} = \frac{1}{t}\left[s_{0} + s_{1}(1  - h(\phi_\mathcal{Z})) - \lambda_{\rm EC} -\lambda_{\rm c} - \lambda_{\rm sec}\right],
 \label{eq:skr}
\end{equation}
where terms $s_{0}$ and $s_{1}$ are the lower bounds on the number of vacuum and single-photon detection events in the key generating $\mathcal{Z}$  basis, $\phi_\mathcal{Z}$ is the upper bound on the phase error rate in the $\mathcal{Z}$ basis corresponding to single-photon pulses, $h(\cdot)$ is the binary entropy, $\lambda_{\rm EC}$ and $\lambda_{\rm c}$ are the number of bits published during the error correction and confirmation of correctness steps, $\lambda_{\rm sec} = 6 \log_2(\frac{19}{\epsilon_{\rm sec}})$ with $\epsilon_{\rm sec}= 10^{-10}$ is the security parameter associated to the secrecy analysis, and finally $t$ is the duration of the quantum transmission phase.
Equation \eqref{eq:skr} is applied to $6.59\cdot 10^6$-bit-long key blocks. This resulted in a finite-key analysis SKR or around $\mathrm{SKR} = 2603 \pm 21$ bits per second whereas the asymptotic SKR is around $\mathrm{SKR}_\infty = 2819 \pm 23$ bits per second.

\section{Conclusions}

In this manuscript, we have proposed a novel QKD source based on a modular design exploiting the iPOGNAC encoder~\cite{Avesani2020} for both intensity and polarization modulation. In this way, our QKD source is immune to side-channel~\cite{Nauerth_2009} and Trojan Horse attacks~\cite{Lee2019} present in sources using multiple lasers, and mitigates intensity pattering effect~\cite{Yoshino2018} without sacrificing the tunability of the decoy state ratio  and while maintaining all benefits deriving from the iPOGNAC. The source was experimentally tested at the NIR optical band around 800 nm, representing the first implementation of the iPOGNAC scheme at this wavelength and confirming the key features of the source. 

The modularity of the scheme is advantageous in the development, testing and qualification of the entire QKD system. This is mainly because a single base element, \textit{i.e.} the iPOGNAC, is responsible for two key tasks in QKD implementation. This allows the system developer to concentrate in optimizing and hardening a single device, without dissipating resources for others.  This is particularly propitious for satellite missions since space-qualification is an expensive and time-consuming process. Furthermore, the design is compatible both at telecom wavelengths and, as demonstrated here, at the NIR optical band, which are  of interest for satellite-based quantum communications. For these reasons, we believe that our work paves the way for the development of a second generation of QKD satellites that can guarantee excellent performances at the highest security levels. 

\begin{acknowledgments}
\noindent Author Contributions: C.A., M.A., G.V., P.V. designed the experiment. A.S., M.A., F.B., C.A. developed the control electronics. F.B., C.A. developed the transmitter and receiver control software and the post-processing software. F.B., S.C. performed the experiment. All authors discussed the results. C.A., F.B. wrote the manuscript with inputs from all the authors.

\noindent This work was supported by the European Union’s Horizon 2020 research and innovation programme, project QUANGO (grant agreement No 101004341) and by MIUR (Italian Minister for Education) under the initiative ``Departments of Excellence'' (Law 232/2016).
\end{acknowledgments}

\bibliographystyle{apsrev4-2}
\bibliography{modularQKD}

\end{document}